\RequirePackage{fix-cm}
\documentclass{svjour3}                     % onecolumn (standard format)
\smartqed  % flush right qed marks, e.g. at end of proof
\usepackage{graphicx}
\usepackage{mathptmx}      % use Times fonts if available on your TeX system
\usepackage{bm}
\newcommand{\Fig}[1]{Fig.\ref{#1}}
\newcommand{\Table}[1]{Table~\ref{#1}}
\newcommand{\Eq}[1]{Eq.~(\ref{#1})}
\begin{document}

\title{
Building diquark model from Lattice QCD
}


\author{Kai Watanabe       \and
        Noriyoshi Ishii %etc.
}


\institute{K.Watanabe \at
              Research Center for Nuclear Physics, Osaka University \\
              Tel.: +81-6-6877-5111\\
              \email{kaiw@rcnp.osaka-u.ac.jp}           %  \\
           \and
           N.Ishii 
           \email{ishiin@rcnp.osaka-u.ac.jp}
}

\date{Received: date / Accepted: date}

\maketitle

\begin{abstract}
A  novel Lattice  QCD  (LQCD) method  to  determine the  quark-diquark
($q$-$D$) interaction potential together with the diquark mass ($m_D$)
is proposed.
Similar  to the  HAL QCD  method, $q$-$D$  potential is  determined by
demanding it to reproduce  the $q$-$D$ equal-time Nambu-Bethe-Salpeter
(NBS) wave function.
%
%% Also, the  masses of quark and  diquark, which are unknown  due to the
%% color confinement, are determined within  our framework.
To do  this, it is necessary  to use the  masses of the quark  and the
diquark as  inputs, which  however are not  straightforwardly obtained
because of the color confinement of QCD.
In this work, masses of quark  and diquark are determined by demanding
that the p-wave spectrums from the two-point correlators be reproduced
by the potentials  for $c\bar{c}$ and $q$-$D$  sectors determined from
the NBS wave functions.
%% We do  so by, first determine  the reduced masses of  the $q$-$\bar q$
%% system  and the  $q$-$D$ system  by  demanding them  to reproduce  the
%% p-wave excitation spectra  for each system and then  combine those two
%% results  and  obtain the  diquark  mass.
%
Numerical calculations  are performed  by using  2+1 flavor  QCD gauge
configurations with the  pion mass $m_\pi\simeq 700$  MeV generated by
PACS-CS collaboration.
We apply our  method to the $c$-$\bar c$ system  and the charm-diquark
system ($\Lambda_c$  baryon) to obtain  the charm quark  mass, diquark
mass and the $c$-$D$ potential.
Our preliminary analysis leads to  the diquark mass $m_D \simeq 1.127$
GeV which  is roughly consistent with  a naive estimate based  on the
constituent quark  picture, i.e., $m_{D} \simeq  m_{\rho} \simeq 1.12$
GeV and $m_{D} \simeq 2m_N/3 \simeq 1.06$ GeV.
%%     to   the   $\rho$   meson   mass
%% $m_\rho\simeq1.12$ GeV and twice  the constituent quark mass $2m_{{\rm
%%         cnst}}\simeq1.06$ GeV.
    %%
\end{abstract}

\section{Introduction}
\label{intro}
Solving quark  many-body problems and revealing  the hadron structures
are important themes in hadron physics.  In general, the computational
complexity  is a  problem  in solving  quantum  many-body systems  and
approximations are often  used.  One example of  such an approximation
is to  reduce the degrees  of freedom  by introducing a  virtual bound
state of the particles constituting the system.  The diquark, which is
a  virtual bound  state  of  two quarks,  is  a  typical example.   By
introducing a  diquark, a baryon  can be expressed as  a quark-diquark
($q$-$D$) bound  state.  A  model based  on the  idea of  diquarks are
called  the  diquark  model  and  has  been  widely  used  to  provide
predictions on hadron structures and energy levels\cite{jaffe}.
 
Since diquarks are color-charged objects,  they cannot be observed due
to the color confinement in the quantum chromodynamics (QCD).  Because
of this  limitation, the parameters in  the diquark model such  as the
$q$-$D$ interaction and  the diquark mass have so  far been determined
on  the  basis  of simplified  phenomenology\cite{mauro}.   Therefore,
determining these  quantities faithfully  to QCD is  indispensable for
the development of hadron physics.
 
 There have been  several studies on diquarks from  lattice QCD (LQCD)
 Monte Carlo  calculation, the  first-principles calculations  of QCD.
 Namely,   Ref.\cite{diquark_mass}   calculated  the   gauge-dependent
 diquark correlation  function and  obtained the  diquark mass  in the
 Landau   gauge.
 %% However,   the  mass   defined  in   this  way   is gauge-dependent.
 To avoid such a  gauge-dependence, Ref.\cite{diquark_size} proposed a
 method to consider  the diquark correlation function  in the presence
 of an infinitely heavy static quark which is introduced to neutralize
 the system.   However, though  this method is  gauge-independent, the
 dynamics  of  the diquark  differ  from  that in  the  experimentally
 observed hadrons due to the introduction of the static quark.

Recently,  a method  to determine  the quark-antiquark  ($q$-$\bar q$)
interaction   potential  from   LQCD   was  proposed   by  Ikeda   and
Iida\cite{Ikeda-Iida}.
In their method, the potential that appears in the interaction term of
the Schr\"odinger equation is determined  by demanding it to reproduce
the equal-time NBS wave function and its energy.
The  $q$-$\bar   q$  potential  calculated   by  Ikeda  and   Iida  is
qualitatively   similar    to   the    phenomenologically   determined
Cornell-type potential\cite{kinoshita}.
However, Ikeda-Iida employed  a very naive estimate of  the quark mass
which is half the vector meson mass.
To  improve   this  point,  Kawanai   and  Sasaki\cite{Kawanai-Sasaki}
proposed  a  method to  determine  the  quark  mass by  demanding  the
spin-spin interaction potential  to vanish at the  long distance.  The
potential and mass determined in  this way reproduce the energy levels
of mesons with satisfactory accuracy\cite{Kawanai-Sasaki2}.
      
These methods of determining the potential  and the mass using the NBS
wave function calculated by LQCD seems to be promising.
However, the Kawanai-Sasaki  method does not work  for quark-diquark
systems where the  diquark is the scalar-diquark  because of the
absence of the spin-spin interaction potential.
Note  that  the scalar-diquark  is  considered  as the  most  relevant
diquark in the hadronic phenomenology.
In  this study,  we propose  an  alternative method  to determine  the
diquark mass.
We determine  the $q$-$D$ potential  by demanding it to  reproduce the
equal-time NBS wave function.
The diquark mass is determined  by demanding this $q$-$D$ potential to
reproduce the p-wave excitation energy of the $q$-$D$ system.
      
This paper is organized as  follows.
The first section  is dedicated  to the  formulation of  our method.
The NBS wave function is defined.  The Schr\"odinger equation is
used to obtain the diquark mass and the $q$-$D$ potential.
To be specific, we focus on the $\Lambda_c$ baryon consisting of
a   spectator  charm   quark  and   a  scalar   [$ud$]  diquark;
$J^{P}=0^+$, isospin $I=0$ and color $\boldmath{\bar{3}}$.
In  the second  section, the  LQCD setup  is explained.
In the third section, we show  the numerical results for the NBS
wave function, the  diquark mass and the  potential.
Finally in the last section, we summarize our work.

\section{Formalism}
\label{sec:1}
We start from the equal-time NBS wave function in the rest frame given by
\begin{equation}
  \psi_{\Lambda(J^P)}(\bm{r})
  \equiv
  \left\langle 0 \left|
  D_c(\bm r )
  c_{c}(\bm 0)
  \right| \Lambda_c(J^P)\right\rangle,
\end{equation}
where $|\Lambda_c(J^{P})\rangle$ denotes  the $\Lambda_c$ baryon state
for $J^{P}$ sector.
$c_c(\bm  x)$   and  $D_c(\bm  x)   \equiv  \varepsilon_{abc}u_a^T(\bm
x)C\gamma_5 d_b(\bm x)$ denote the field operators for the charm quark
and the composite diquark, respectively.
$u_a(\bm x)$, $d_a(\bm  x)$ denote the field operators for  the up and
the down quarks,  respectively, with $C$ being  the charge conjugation
matrix.
$a$,$b$ and $c$ are the color indices.
The Levi-Civita  symbol $\varepsilon_{abc}$ is introduced  to make the
color representations of the diquark operator to be $\bar{\bm 3}$.
To obtain  the NBS wave  function, we consider the  $c$-$D$ four-point
correlator $ C(\bm x  - \bm y, t) \equiv \langle 0|  T\{ D_c(\bm x, t)
c_c(\bm y,  t)\cdot {\mathcal  J}_{\Lambda(J^P)}(t=0) \} |  0\rangle $
with $\mathcal{J}_{\Lambda(J^P)}$  being the wall source  operator for
the $\Lambda(J^P)$ baryon in our calculation.
%%%
The NBS wave function is obtained in the large $t$ region of $C(\bm r,
t)$  as  $ C(\bm  r,  t)  \simeq  \psi_{\Lambda(J^P)}(\bm r)  \cdot  A
\exp\left(  -M_{\Lambda(J^P)}  t  \right)$   with  $A  \equiv  \langle
\Lambda_{J^P}| {\mathcal J}_{J^P} | 0\rangle$.

We define the  quark-diquark potential $\hat U$ by  demanding that the
following Schr\"odinger  equation be  satisfied by the  equal-time NBS
wave function as
\begin{equation}
  \left(
  -\frac{\nabla^2}{2\mu_{cD}}
  +\hat{U}
  \right)
  \psi_{\Lambda(J^P)}(\bm r)
  =
  E_{\Lambda(J^{P})}
  \psi_{\Lambda(J^{P})}(\bm r),
  \label{eq:schrodinger_qD0}
\end{equation}
where $\mu_{cD}\equiv  m_cm_D/(m_c+m_D)$ denotes  the reduced  mass of
the system  with $m_c$ and $m_D$  being the masses of  charm quark and
the diquark respectively.
We  treat   these  masses  as   unknown  parameters  at   this  point,
which will be determined later.
$E_{\Lambda(J^{P})}\equiv M_{\Lambda(J^{P})} -  (m_c+m_D)$ denotes the
binding energy with $M_{\Lambda(J^{P})}$ being the mass of $\Lambda_c$
baryon for $J^P$ channel.
We apply the derivative expansion to $\hat  U$ as
\begin{equation}
  \hat{U}
  =
  U_{0}(r)
  +
  U_{\rm LS}(r)\bm L \cdot \bm s  + O(\nabla^2),
 \label{eq:derivative_expansion_qD}
\end{equation}
where  $U_{0}(r)$, $U_{\rm  LS}(r)$, $\bm  L$ and  $\bm s$  denote the
central and  the spin-orbit  potentials, the orbital  angular momentum
operator and the spin operator of the charm quark, respectively.
To proceed, we define the following quantity:
\begin{equation}
  \widetilde U(\bm r)
  \equiv
  \frac{\nabla^2 \psi_{\Lambda(1/2^+)}(\bm r)}{\psi_{\Lambda(1/2^+)}(\bm r)}
  \simeq
  2\mu_{cD} \left(U_0(\bm r) - E_{\Lambda(1/2^+)}\right),
  \label{eq:pre-pot}
\end{equation}
which we refer to as the {\it pre-potential}.
Due  to \Eq{eq:schrodinger_qD0},  NBS wave  functions for  $J^P=1/2^-,
3/2^-$ satisfy the following {\it pre-Schr\"odinger equation}:
\begin{equation}
  \left(
  - \nabla^2
  +
  \widetilde U(\bm r)
  +
  2\mu_{cD} U_{\rm LS}(\bm r)\bm L\cdot\bm s
  \right)
  \psi_{\Lambda(J^P)}(\bm r)
  %%%
  =
  %%%
  2\mu_{cD}
  \left(
  M_{\Lambda(J^P)} - M_{\Lambda(1/2^+)}
  \right)
  \psi_{\Lambda(J^P)}(\bm r).
\end{equation}
%%%
By treating the  spin-orbit potential $U_{\rm LS}(r)\bm  L\cdot \bm s$
as a perturbation, the pre-Schr\"odinger  equations are solved in $J^P
= 1/2^-$ and $3/2^-$ sectors to have
\begin{eqnarray}
  \widetilde E_{\rm PW}
  -2\mu_{cD}\left\langle U_{\rm LS}\right\rangle
  &\simeq&
  2\mu_{cD}\left(M_{\Lambda(1/2-)} - M_{\Lambda(1/2+)}\right)
  \\\nonumber
  \widetilde E_{\rm PW}
  + \mu_{cD}\left\langle U_{\rm LS}\right\rangle
  &\simeq&
  2\mu_{cD}\left(M_{\Lambda(3/2-)} - M_{\Lambda(1/2+)}\right),
\end{eqnarray}
where $\widetilde E_{\rm PW}$  denotes the degenerate unperturbed {\it
  pre-energy} for the p-wave sector  which is obtained by solving the
unperturbed {\it pre-Schr\"odinger equation}
\begin{equation}
  \left(
  -\nabla^2 + \widetilde U(\rm r)
  \right)
  \psi(\bm r)
  =
  \widetilde E
  \psi(\bm r),
  \label{eq:unperturbed_pre_schrodinger}
\end{equation}
in the p-wave sector.
%%
%% By eliminating $\langle U_{\rm LS}\rangle$ from these equations and by
%% using   $E_{\Lambda(J^P)}$  obtained   from  the   temporal  two-point
%% correlators, we arrive at the relation
By eliminating  $\langle U_{\rm LS}\rangle$ from  these two equations,
we have
\begin{equation}
  \mu_{cD}
  =
  \frac{3}{2}\frac{\widetilde E_{\rm PW}}{M_{\Lambda(1/2-)} + 2M_{\Lambda(3/2-)} - 3M_{\Lambda(1/2+)}},
  \label{eq:mu_cd}
\end{equation}
which   enables  us   to   determine  the   reduced   mass  by   using
$M_{\Lambda(J^P)}$ obtained from the two-point correlators.
The diquark mass $m_{D}$ can be  obtained by combining the charm quark
mass  $m_c$ which  can be  obtained by  applying a  similar method  to
$c\bar{c}$ sector.
Finally,   the   quark-diquark   potential  is   obtained   from   the
pre-potential as
\begin{equation}
  U_0(\bm r)
  =
  \frac1{2\mu_{cD}} \widetilde U_{\Lambda(1/2^+)}(\bm r)
  + M_{\Lambda(1/2^+)} - m_{D} - m_{c}.
  \label{eq:final_potential}
\end{equation}

\section{LQCD setup}
\label{setup}
In this work, we use 2+1  flavor QCD gauge configurations generated by
PACS-CS Collaboration on $32^3  \times 64$ lattice \cite{pacs_config},
which employs  the RG improved Iwasaki  gauge action at $\beta  = 1.9$
\cite{iwasaki1}  and  the  non-perturbatively $O(a)$  improved  Wilson
quark  action  at  $(\kappa_{\rm  ud},  \kappa_{\rm  s})  =  (0.13700,
0.13640)$ and $C_{\rm SW}=1.715$ \cite{cl_wilson}.
This parameter  set leads to the  lattice spacing $a =  0.0907(13)$ fm
($a^{-1} \simeq 2.175$ GeV), the spatial extent $L=32a \simeq 2.9$ fm,
the pion  mass $m_{\pi} \simeq  700$ MeV  and the nucleon  mass $m_{N}
\simeq 1584$ MeV.
For the charm quark, the relativistic heavy quark action (RHQ) is used
in order  to reduce the  systematic errors originating from  the heavy
charm quark mass \cite{RHQ_on_LQCD}.
We use the same parameter set as given in Ref.\cite{namekawa}.
The  lattice  QCD  calculation  is  carried out  by  using  399  gauge
configurations  for   several  different  source  points   for  better
statistics.
The  statistical   errors  are   evaluated  by  using   the  Jackknife
prescription.
We employ the  Coulomb  gauge  fixing though out the calculations.

\section{Numerical results}
\label{num}
Fig.1 shows the NBS wave function for $\Lambda_c(1/2^+)$ extracted from
the   $c$-$D$   four-point    correlation   function   at   time-slice
$t/a=15$.  Note that  $t/a =  15$ is  the smallest  time-slice in  the
plateau region.
%We referred to \Fig{fig:1} but gives (Fig.4, not Fig.1)
\begin{figure*}[h]
\begin{center}
  \includegraphics[width=0.50\textwidth,natwidth=600,natheight=600]{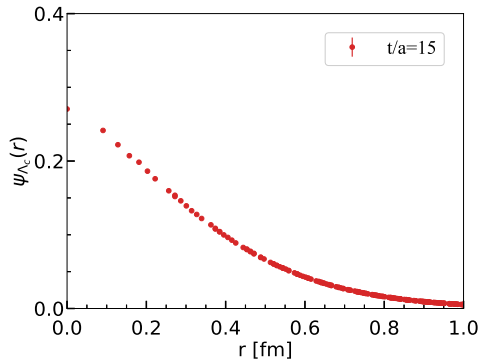}
\caption{$c$-$D$ NBS wave function of the $\Lambda_c(\frac{1}{2}^+)$ state. 
	The NBS wave function is extracted from the four-point function at $t/a=15$.
	The NBS wave function is normalized by using $L^2$-norm.}
\end{center}
\label{fig:1}       
\end{figure*}
\Fig{fig:2}  shows the  result  of  the pre-potential  \Eq{eq:pre-pot}
together with the result of the fit with the Cornell-plus-log type fit
function
%% The result for the pre-potential \Eq{eq:pre-pot} is shown in \Fig{fig:2}.
%% The result of the Cornell-plus-log type function(
\begin{equation}
 V^{{\rm fit}}(r) 
 =
- A/r
+
B r
+
C\log(r/a)
+
{\rm const}.
\end{equation}
%
%
%% Here,  the  log term  introduced  to  correct  the finite  quark  mass
%% effects\cite{Koma_Koma}.
The log term is introduced to improve the quality of the fit. In the $q$-$\bar q$
sector,  it is  suggested that  such log  term may  appear due  to the
finite quark mass effect \cite{Koma_Koma}.
To avoid an  artifact from the spatial boundary  condition, we perform
the fit in a restricted region $0 < r/a \leq 10$.
\begin{figure*}% For one-column wide figures use
\begin{center}
  \includegraphics[width=0.50\textwidth]{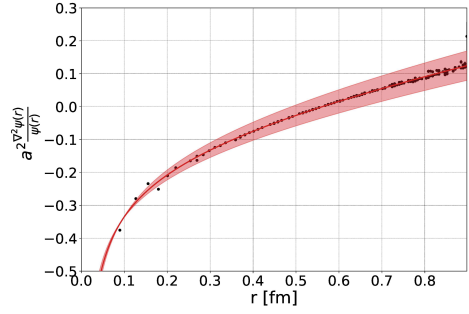}
\caption{$c$-$D$ pre-potential (black dots) and  the result of the fit
  with  the  Cornell-plus-log  function(red line).   The  shaded  area
  denotes the statistical error.}
	\label{fig:2}    
	\end{center}
\end{figure*}

\begin{figure*}% For one-column wide figures use
\begin{center}
  \includegraphics[width=0.50\textwidth]{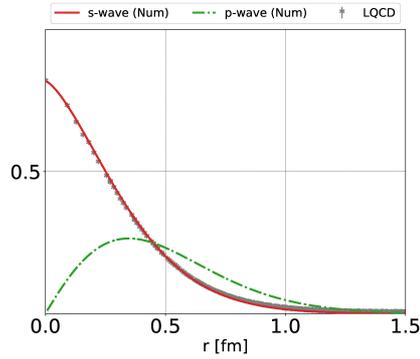}
  \caption{Numerical solutions  for s-wave (red solid  line) and
    p-wave (green dotted line)  obtained from the pre-potential.
    The LQCD data (star) is  also shown for comparison.  The NBS
    wave function is normalized accordingly.}
  \label{fig:3}    
\end{center}
\end{figure*}

Next,    we     use    the     fitted    pre-potential     to    solve
\Eq{eq:unperturbed_pre_schrodinger}     to    obtain     the    p-wave
excitation pre-energy $\widetilde E_{\rm PW}$.
To solve the equation numerically,  we employ the Discretized Variable
Representation   (DVR)  method   often  used   in  quantum   chemistry
\cite{DVR1}.
We obtain $\widetilde E_{\rm PW} = 0.641(4)$ GeV$^2$.
The results for s-wave and  p-wave solutions are shown in \Fig{fig:3}.
We  see a  good  agreement between  the LQCD  data  and the  numerical
solution for the s-wave.

We  calculate   the  two-point   correlators  to  obtain   the  masses
$M_{\Lambda(1/2^+)}$, $M_{\Lambda(1/2^-)}$ and $M_{\Lambda(3/2^-)}$ of
$\Lambda_c$ baryons.  The results are summarized in \Table{tab:1}.
%%
%% The p-wave  unperturbed excitation  energy for  the $c$-$D$  system is
%% $E_{PW}     \simeq      \frac{1}{3}\left(M_{\Lambda(1/2-)}     +     2
%% M_{\Lambda(3/2-)}\right) - M_{\Lambda(1/2+)} \simeq 460$ MeV.
\begin{table}
  \begin{center}
    \caption{The energy levels of $\Lambda_c$ baryons.}
    \label{tab:1}       % Give a unique label
    \begin{tabular}{ccc}
      \hline\noalign{\smallskip}
      $\frac{1}{2}^+$ & $\frac{1}{2}^-$ & $\frac{3}{2}^-$  \\
      \noalign{\smallskip}\hline\noalign{\smallskip}
      2.684(4) GeV& 3.083(81) GeV & 3.176(12) GeV\\
      \noalign{\smallskip}\hline
      %from 2020/0126 data
    \end{tabular}
  \end{center}
\end{table}
The reduced mass $\mu_{cD}$ is now calculated by using \Eq{eq:mu_cd}.
%% Making use  of the excitation  pre-energy obtained above,  the reduced
%% mass $\mu_{cD}$ is determined from \Eq{eq:mu_cd}.
%%
Combining  $\mu_{cD}$ and  the  charm quark  mass $m_c=1.823(19)$  GeV
obtained by applying the similar procedure to the $c$-$\bar c$ sector,
the diquark mass is obtained as $m_D=1.127(114)$ GeV.
Now the $c$-$D$ potential is obtained by using \Eq{eq:final_potential}.
Note that our diquark mass is roughly consistent with a naive estimate
based  on the  constituent  quark picture,  i.e.,  $m_D \simeq  m_\rho
\simeq 1.12$ GeV and $m_D \simeq 2 m_N/3 \simeq 1.06$ GeV.
We make a  comment. In Fig.1 and Fig.2, we  recognize that the lattice
QCD data of the NBS wave function and the pre-potential deviate from a
single-valued function of $r$ at short distance,
which  seems to  be  due  to the  discretization  artifact of  lattice
Coulomb gauge.
As a result,  the fit of the pre-potential at  short distance receives
an uncertainty,  which may lead to  a systematic uncertainty of  a few
hundred MeV  of the diquark  mass. There is  a similar problem  in the
$q\bar{q}$ sector, which  may lead to an uncertainty of  a few hundred
MeV in the determination of the charm quark mass.
To improve, careful analysis is needed.
%
%% Once the diquark mass is  determined, the potential is simply obtained
%% by  dividing  the  pre-potential   \Fig{fig:3}  by  the  reduced  mass
%% $2\mu_{cD}$      followed     by      a     constant      shift     by
%% $E_{\Lambda(\frac{1}{2}^+)}$.

%%
%which lies close  to the $\rho$
%meson mass  $m_\rho\simeq1.12$ GeV and double  the constituent quark
%mass  $2m_{{\rm cnst}}\simeq1.06$  GeV.   Once the  diquark mass  is
%determined,  the   potential  is  simply  obtained   by  dividing  the
%pre-potential \Fig{fig:3} by the mass.
\section{Sammary}
\label{sec:5}
We have  proposed a novel  method to obtain  the diquark mass  and the
quark-diquark interaction potential from LQCD.
By  using 2+1  flavor QCD  gauge configurations  generated by  PACS-CS
collaboration  with $m_{\pi}  \simeq  700$ MeV,  we  have performed  a
lattice QCD Monte Carlo calculation.
A preliminary analysis has lead to the diquark mass $m_D = 1.127(114)$
GeV,  which  is  consistent  with   a  naive  estimate  based  on  the
constituent quark picture, i.e., $m_D \simeq m_{\rho} \simeq 1.12$ GeV
and $m_D \simeq 2m_N/3 \simeq 1.06$ GeV.
We  have obtained  a  quark-diquark potential  which is  qualitatively
similar to the well-known Cornell type potential.

\begin{acknowledgements}
We thank  Y.~Ikeda, S.~Watanabe, M.~Koma and  A.~Nakamura for fruitful
discussions.
The  Lattice  QCD  calculation  has  been carried  out  by  using  the
supercomputer OCTOPUS at Cyber Media  Center of Osaka University under
the  support  of   Research  Center  for  Nuclear   Physics  of  Osaka
University.
We thank PACS-CS Collaboration and ILDG/JLDG for providing us with the
2+1          flavor          QCD         gauge          configurations
\cite{pacs_config,beckett,ILDG,JLDG}.  The lattice QCD code is partly based on Bridge++\cite{Bridge} 
This  research is  supported by MEXT  as  ``Program  for  Promoting Researches  on  the  Supercomputer
Fugaku''  (Simulation  for basic  science:  from  fundamental laws  of
particles to creation  of nuclei) and JICFuS.  This  work is supported
by JSPS KAKENHI Grant Number JP21K03535.
\end{acknowledgements}

\bibliographystyle{spphys}       % APS-like style for physics
\bibliography{biblist}   % name your BibTeX data base

\begin{thebibliography}{10}
\providecommand{\url}[1]{{#1}}
\providecommand{\urlprefix}{URL }
\expandafter\ifx\csname urlstyle\endcsname\relax
  \providecommand{\doi}[1]{DOI \discretionary{}{}{}#1}\else
  \providecommand{\doi}{DOI \discretionary{}{}{}\begingroup
  \urlstyle{rm}\Url}\fi

\bibitem{jaffe}
R.~Jaffe, Physics Reports \textbf{409}(1), 1  (2005)

\bibitem{mauro}
M.~Anselmino, E.~Predazzi, S.~Ekelin, S.~Fredriksson, D.B. Lichtenberg, Rev.
  Mod. Phys. \textbf{65}, 1199 (1993)

\bibitem{diquark_mass}
M.~Hess, F.~Karsch, E.~Laermann, I.~Wetzorke, Phys. Rev. D \textbf{58}, 111502
  (1998)

\bibitem{diquark_size}
C.~Alexandrou, P.~de~Forcrand, B.~Lucini, Phys. Rev. Lett. \textbf{97}, 222002
  (2006)

\bibitem{Ikeda-Iida}
Y.~Ikeda, H.~Iida, Progress of Theoretical Physics \textbf{128}(5), 941 (2012)

\bibitem{kinoshita}
E.~Eichten, et~al., Phys. Rev. Lett. \textbf{34}, 369 (1975)

\bibitem{Kawanai-Sasaki}
T.~Kawanai, S.~Sasaki, Phys. Rev. Lett. \textbf{107}, 091601 (2011)

\bibitem{Kawanai-Sasaki2}
T.~Kawanai, S.~Sasaki, Phys. Rev. D \textbf{92}, 094503 (2015)

\bibitem{pacs_config}
S.~Aoki, et~al., Phys. Rev. D \textbf{79}, 034503 (2009)

\bibitem{iwasaki1}
Y.~Iwasaki.
\newblock unpublished (2011)

\bibitem{cl_wilson}
S.~Aoki, M.~Fukugita, S.~Hashimoto, K.I. Ishikawa, N.~Ishizuka, Y.~Iwasaki,
  K.~Kanaya, T.~Kaneko, Y.~Kuramashi, M.~Okawa, S.~Takeda, Y.~Taniguchi,
  N.~Tsutsui, A.~Ukawa, N.~Yamada, T.~Yoshi\'e, Phys. Rev. D \textbf{73},
  034501 (2006)

\bibitem{RHQ_on_LQCD}
S.~Aoki, et~al., Progress of Theoretical Physics \textbf{109}(3), 383 (2003)

\bibitem{namekawa}
Y.~Namekawa, et~al., Phys. Rev. D \textbf{84}, 074505 (2011)

\bibitem{Koma_Koma}
Y.~Koma, M.~Koma, Few-Body Systems \textbf{54} (2013)

\bibitem{DVR1}
D.T. Colbert, W.H. Miller, The Journal of Chemical Physics \textbf{96}(3), 1982
  (1992)

\bibitem{beckett}
M.G. Beckett, P.~Coddington, B.~Jo^^c3^^b3, C.M. Maynard, D.~Pleiter,
  O.~Tatebe, T.~Yoshie, Computer Physics Communications \textbf{182}(6), 1208
  (2011)

\bibitem{ILDG}
{International Lattice Data Grid (ILDG)}.
\newblock \url{http://www.lqcd.org/ildg}

\bibitem{JLDG}
{Japan Lattice Data Grid (JLDG)}.
\newblock \url{http://www.jldg.org}

\bibitem{Bridge}
{Lattice QCD code Bridge++}.
\newblock \url{http://bridge.kek.jp/Lattice-code/}

\end{thebibliography}

%\begin{thebibliography}{}
% etc
%\end{thebibliography}

\end{document}